\documentclass{pasj00}
\draft

\begin{document}
\SetRunningHead{}{Short-Term Variability of PKS~1510$-$089}

\title{An Intrinsic Short-term Radio Variability Observed in PKS~1510$-$089}

\author{%
  Akiko \textsc{Kadota}\altaffilmark{1},
  Kenta \textsc{Fujisawa}\altaffilmark{1,2},
  Satoko \textsc{Sawada-Satoh}\altaffilmark{3},
  Kiyoaki \textsc{Wajima}\altaffilmark{2},
  and
  Akihiro \textsc{Doi}\altaffilmark{4}}
\altaffiltext{1}{The Research Institute for Time Studies, Yamaguchi University, Yoshida 1677-1, Yamaguchi-city, Yamaguchi 753-8511}
\altaffiltext{2}{Department of Physics, Faculty of Science, Yamaguchi University, Yoshida 1677-1, Yamaguchi-city, Yamaguchi 753-8512}
\altaffiltext{3}{Mizusawa VLBI Observatory, National Astronomical Observatory of Japan, Hoshigaoka-cho 2-12, Oshu, Iwate 023-0861}
\altaffiltext{4}{Institute of Space and Astronautical Science, Japan Aerospace Exploration Agency, Yoshinodai 3-1-1, Chuo-ku, Sagamihara, Kanagawa 252-5210}
\email{kadota@yamaguchi-u.ac.jp}

%

\KeyWords{galaxies: active --- galaxies: quasars: individual (PKS~1510$-$089) ---  radio continuum: galaxies} 
\maketitle

\begin{abstract}
We searched a short-term radio variability in an active galactic nucleus PKS~1510$-$089.
A daily flux monitoring for 143 days at 8.4~GHz was performed,
and VLBI observations at 8.4, 22, and 43~GHz were carried out 4 times during the flux monitoring period.
As a result, variability with time scale of 20 to 30 days was detected.
The variation patterns were well alike on three frequencies, moreover those at 22 and 43~GHz were synchronized.
These properties support that this short-term variability is an intrinsic one.
The Doppler factor estimated from the variability time scale is 47.
Since the Doppler factor is not extraordinary large for AGN,
such intrinsic variability with time scale less than 30 days would exist in other AGNs.
\end{abstract}

\section{Introduction}

It is known that radio flux of active galactic nuclei (AGNs) vary for various time scales (e.g., \cite{1985ApJS...59..513A}; \cite{2005A&A...440..409T}).
According to \citet{1992ApJ...396..469H}, time scale of AGN flare is two years on the average. 
The Intra-day Variability (IDV), a flux variability of AGN
with time scale of about one day is also known (\cite{1992A&A...258..279Q}, \cite{2010ARep...54..908G}).
This very fast variability is, however,
probably caused by interstellar scintillation \citep{2000aprs.conf..147J}, and not an intrinsic one.
At mm radio band an intrinsic variability was reported for S5~0716+71 with a timescale shorter than 30 days at 86, 229~GHz
(\cite{2006A&A...456..117A}, \cite{2006A&A...451..797O}).

At high energy band, for example optical, X-ray, and gamma-ray, it is known that the time scale
of flux variability is much shorter than a month because such photons emanate from small region with a size of less than one light-month.
The optical depth at microwave band is thought to be larger than unity for such small region,
and the radio emission from such small region cannot be observed.
The observed region ($\tau = 1$ surface) is relatively large,
and the time scale of intrinsic variability is thought to be longer than 30 days at microwave.
In addition, the variability of microwave band shows low correlation with the variability at high-energy bands.

However, search for such short-term variability have not yet be made intensively
(one important exception is the ongoing OVRO 15 GHz program \citep{2011ApJS..194...29R}, which is monitoring over 1000 AGNs with observations made every few days).
Even at low frequency below 10~GHz, such short-term variation was reported by the continuous monitoring observation (\cite{2001ApJS..136..265L}).
For example, variability with amplitude of 0.2~Jy at 8~GHz and timescale of 20 to 30 days is
visible in the BL Lac object 1749+096 by the daily monitoring of the Green Bank Interferometer \citep{2001ApJS..136..265L}. 

It is still not clear if such variability would be an extrinsic one similar to IDV, or intrinsic one.
So we focus on the spectral variability.
The suspected intrinsic variability would be correlatively observed at different frequencies like large radio flares \citep{1985ApJS...59..513A}.
If we find a correlated variability at different frequencies, it would be a piece of evidence that the variability is intrinsic.
In order to search a short-term intrinsic variability, we have made a daily observation and a multi-frequency VLBI observation.

In this paper, we describe first the selected target PKS~1510$-$089 in section 2,
and the daily flux monitoring at 8.4~GHz
and the multi-frequency VLBI observation for four times in section 3.
In section 4, we present the results, and we discuss about the observed short-term variability in section 5. 
In this paper, we assume flat universe with $H_0=71.0$~km~s$^{-1}$~Mpc$^{-1}$ and $\Omega _{\mathrm{M}}=0.270$.

\section{PKS~1510$-$089}
PKS~1510$-$089 ($z=0.360$; \cite{1990PASP..102.1235T}, 1~mas = 5.0~pc) is one of the most extreme AGNs showing
strong and rapid variability at all spectral range, and it would be a suitable source for searching a short-term variability.
In the optical band, this object is classified as highly polarized quasar  \citep{1967ApJ...149L..17A}.
\citet{2010ApJ...710L.126M} and \citet{2011PASJ...63..489S} reported that
rotation of optical polarization vector for 50 days when a prominent optical
flare were observed.
This object exhibits strong X-ray emission and short-term variability (e.g., \cite{2003A&A...401..505G}; \cite{2010ApJ...710L.126M}).
\citet{2010ApJ...710L.126M} reported that the X-ray and radio fluxes were significantly correlated,
although the variability timescale seemed to be longer than 30 days.
GeV gamma-ray was also detected by CGRO/EGRET \citep{1999ApJS..123...79H},
and gamma-ray variability were recently observed by Fermi-LAT and
AGILE on timescales of $6-12$~hr (\cite{2010ApJ...721.1425A}; \cite{2009A&A...508..181D}; \cite{2011A&A...529A.145D}).
In the radio band, light curves of total flux density observed by University of Michigan Radio Astronomy Observatory
\citep{1985ApJS...59..513A} and Mets{\"a}hovi Radio Observatory \citep{2005A&A...440..409T} shows intense variability
with time scale less than one year.
A number of VLBI observations have been performed, and it is revealed that the emission
is strongly core dominated and jet components are seen in the northwest direction
from the core (\cite{2004ApJS..150..187O}; \cite{2005AJ....130.1418J}; \cite{2009AJ....137.3718L}).
Apparent velocity of the jets exceeded $20c$ (\cite{2009AJ....138.1874L}).
\citet{2005AJ....130.1418J} has reported an existence of highly relativistic jet which exhibit
superluminal motion with velocity of $\sim46c$.
Viewing angle of the jet is estimated as $\sim 2.5^\circ $
and jet is almost straight toward the observer. Flux variability of PKS~1510$-$089
is also observed by VLBI monitoring observation.
In VLBA observation at 15~GHz and 22~GHz for one year at two month intervals by \citet{2004ApJS..150..187O},
flux variability was observed in the core and jet components for every epoch.
In VLBA observation at 43~GHz at two month intervals by \citet{2005AJ....130.1418J},
it is reported that there is relation at the time of the jet ejection and flare of total flux density.

\section{Observation and Data Reduction}
\subsection{Total flux density monitoring}
The flux density monitoring was carried out with Yamaguchi 32~m radio telescope for
the period of 143 days from 2010 February 1 to 2010 June 24
(corresponding to DOY 32 to 175; DOY is the day of year 2010. DOY$=0$ of 2010 corresponds to MJD 55196). 
We measured the radio emission at 8.4~GHz with a bandwidth of 400~MHz in the total power mode.
The full-width at half maximum (FWHM) of the beam is 4.2~arcmin.
Since there is no beam switching system in Yamaguchi 32~m, we adopted 'Z-scan' method to measure the flux density
by removing gain and atmospheric fluctuation and pointing offsets.
Data are continuously acquired during that the telescope moves four points around the target with offset of 
($-20^\prime $, $-2^\prime $), ($+20^\prime $, $-2^\prime $), ($-20^\prime $, $+2^\prime $), ($+20^\prime $, $+2^\prime $) in right ascension and declination, respectively.
The scan pattern likes flat 'Z' letter.
The observed power of each scan was fitted with a Gaussian function, then the true peak power was calculated from the four scans by removing the declination offset.
A strong, nearby radio source 1453$-$109 was used as a flux density calibrator. The angular size of this source is smaller than the beam size of Yamaguchi 32~m,
and the flux density of this source is well stable. The flux density of 1453$-$109 at 8.4~GHz was measured as 0.93~Jy by Yamaguchi 32~m on 2010 May 1,
by referencing the flux density calibrator 3C~286 (5.21~Jy, \cite{1994A&A...284..331O}). 
The calibrator 1453$-$109 and the target PKS~1510$-$089 were observed alternately for 21 minutes each at the same elevation range,
and the flux density of PKS~1510$-$089 is relatively determined with the flux density of 1453$-$109.
Observations were made at dual circular polarization simultaneously.
The standard deviation of the measurement was typically 0.06~Jy.
Due to the weather condition, effective observation was made at 72 times.
The observation was not made from 2010 May 17 to 2010 June 2 (DOY 137 to 153) due to another observation.

\subsection{VLBI Observations}
VLBI observations at 8.4, 22, and 43~GHz were carried out four times at each frequency every about ten days from March 26 to May 1, 2010.
Four telescopes of VLBI Exploration of Radio Astrometry (VERA) were used for 22 and 43~GHz observations. At 8.4~GHz we used 4 or 5 telescopes (Yamaguchi 32~m, Kashima 34~m, and VERA) of the Japanese VLBI Network (JVN).
Observations by 8.4, 22, and 43~GHz were performed in consecutive three days.
The 3rd and 4th observation (April 22 and May 1) at 43~GHz were not made in consecutive because of the receiver trouble at Iriki station.
The dates and stations of the VLBI observation are listed in table \ref{tabl:table1}.
For 8.4~GHz observation, right-hand circular polarization were received in two IF bands with 16~MHz bandwidth.
The data was quantized at 2-bit, and recorded on magnetic tape at a rate of 128~Mbps.
At 22~GHz and 43~GHz, left-hand circular polarization was received in bandwidth of 128~MHz by dual-beam mode of VERA.
The data was quantized at 2-bit, and recorded on magnetic tape at a rate of 1024~Mbps. A single beam data (512~Mbps) was used for this study.
For all data, correlation processing was performed using Mitaka FX correlator.
The total on-source time of PKS~1510$-$089 were 5 and 5.9 hours for JVN and VERA observations, respectively.
3C~345 and NRAO~512 were observed for bandpass and amplitude calibration.

The data reduction was performed with the Astronomical Image Processing System (AIPS) of NRAO.
At 22~GHz and 43~GHz, a standard a priori amplitude calibration was carried out by using the AIPS task APCAL.
At 8.4~GHz, a standard a priori amplitude calibration was not used, 
because some JVN telescopes do not have the $T_\mathrm{sys}$ monitoring system, and gain curves of some telescopes were not well known.
The scaling factor of amplitude was obtained from the comparison with correlated and total flux density of the flux calibrator NRAO~512.
The total flux density of NRAO~512 was measured with Yamaguchi 32~m on 2010 May 1 after the VLBI monitoring observation, and it was 1.27~Jy.
We applied scaling factor of NRAO~512 which almost same time and almost same elevation with PKS~1510$-$089 to all on-source time.
In the amplitude calibration process of epoch 4 at 8.4~GHz, 
Mizusawa station has been excluded since the amplitude gain factor was extraordinary larger than the other three epochs.
We estimated that the random errors in amplitude calibration were 3 to 6~\% for 22 and 43~GHz.
There seems to be other unknown offsets in amplitude, and the total errors were about 10~\%.
Note that the offsets would not cause the false variability.
The estimated error was 8 to 11~\% for 8.4 GHz.
The processes after the amplitude calibration were the same at all bands.
Fringe fitting was performed by the AIPS task FRING, and bandpass calibration was performed by the task BPASS. 
After that, we iterated CLEAN and self-calibration using the Difmap software package and obtained the CLEAN image.

\begin{table*}[tbp]
\caption{VLBI Observations.}
\label{tabl:table1}
\begin{center}
\begin{tabular}{ccrll} \hline
Freq. & Epoch & DOY & Date and Time & Telescopes \\
(GHz)  &         &   & \\\hline
8.4 & X1 & 85 & 2010 Mar 26 13:45-22:15 & M, R, I, Y\\
    & X2 & 98 & 2010 Apr 08 12:55-21:25 & M, R, O, I, Y\\
    & X3 & 108 & 2010 Apr 18 12:15-20:45 & M, R, O, I, K\\
    & X4 & 118 & 2010 Apr 28 11:35-20:05 & M, R, O, I, K\\\hline
22  & K1 & 86 & 2010 Mar 27 13:45-22:15 & M, R, O, I\\
    & K2 & 96 & 2010 Apr 06 13:05-21:35 & M, R, O, I\\
    & K3 & 106 & 2010 Apr 16 12:25-20:55 & M, R, O, I\\
    & K4 & 116 & 2010 Apr 26 11:45-20:15 & M, R, O, I\\\hline
43  & Q1 & 87 & 2010 Mar 28 13:45-22:15 & M, R, O, I\\
    & Q2 & 97 & 2010 Apr 07 13:00-21:30 & M, R, O, I\\
    & Q3 & 112 & 2010 Apr 22 12:00-20:30 & M, R, O, I\\
    & Q4 & 121 & 2010 May 01 11:25-19:55 & M, R, O, I\\\hline
 \multicolumn{5}{@{}l@{}}{\hbox to 0pt{\parbox{110mm}{\footnotesize
Notes. 
Col. 1: observation frequency; 
Col. 2: observation epoch; 
Col. 3: Day of year 2010; 
Col. 4: observation date and time (UTC); 
Col. 5: Stations. M=Mizusawa, R=Iriki, O=Ogasawara, I=Ishigaki, Y=Yamaguchi, K=Kashima.
 }\hss}}
\end{tabular}
\end{center}
\end{table*}

\section{Results}

\subsection{Total flux density at 8.4~GHz}

\begin{figure}
  \begin{center}
    \FigureFile(80mm,80mm){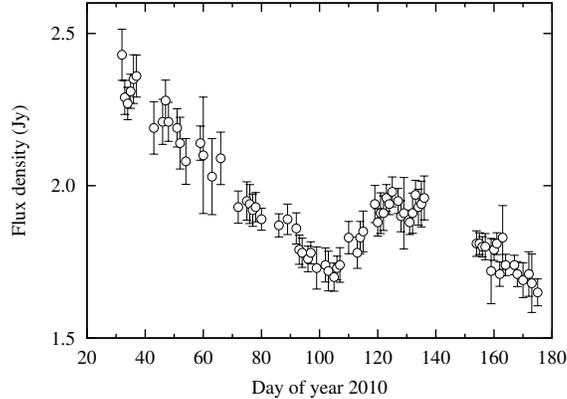}
  \end{center}
  \caption{The light curve of total flux density of PKS~1510-089 at 8.4~GHz. The error bar represents the standard deviation of the measurement of each day.}
  \label{fig:figure1}
\end{figure}

Figure \ref{fig:figure1} shows the light curve of the total flux density of PKS~1510$-$089 at 8.4~GHz.
Flux density monotonically decreased from the beginning (DOY 32) and reached the minimum around DOY 100.
Then the flux density increased and reached $\sim 2$~Jy around DOY 130.
After that, it decrease again, and the flux density was 1.65~Jy at the end of the observation (DOY 175).
In order to estimate the date and flux density of the minimum and the maximum, 
we fit a third order polynomial to the light curve from DOY 86 to DOY 136 (dotted line in figure \ref{fig:figure4}). 
The flux density of minimum was 1.74~Jy in DOY 100, and the maximum was 1.95~Jy in DOY 130, respectively.

\subsection{VLBI images}
\begin{figure*}
  \begin{center}
    \FigureFile(160mm,80mm){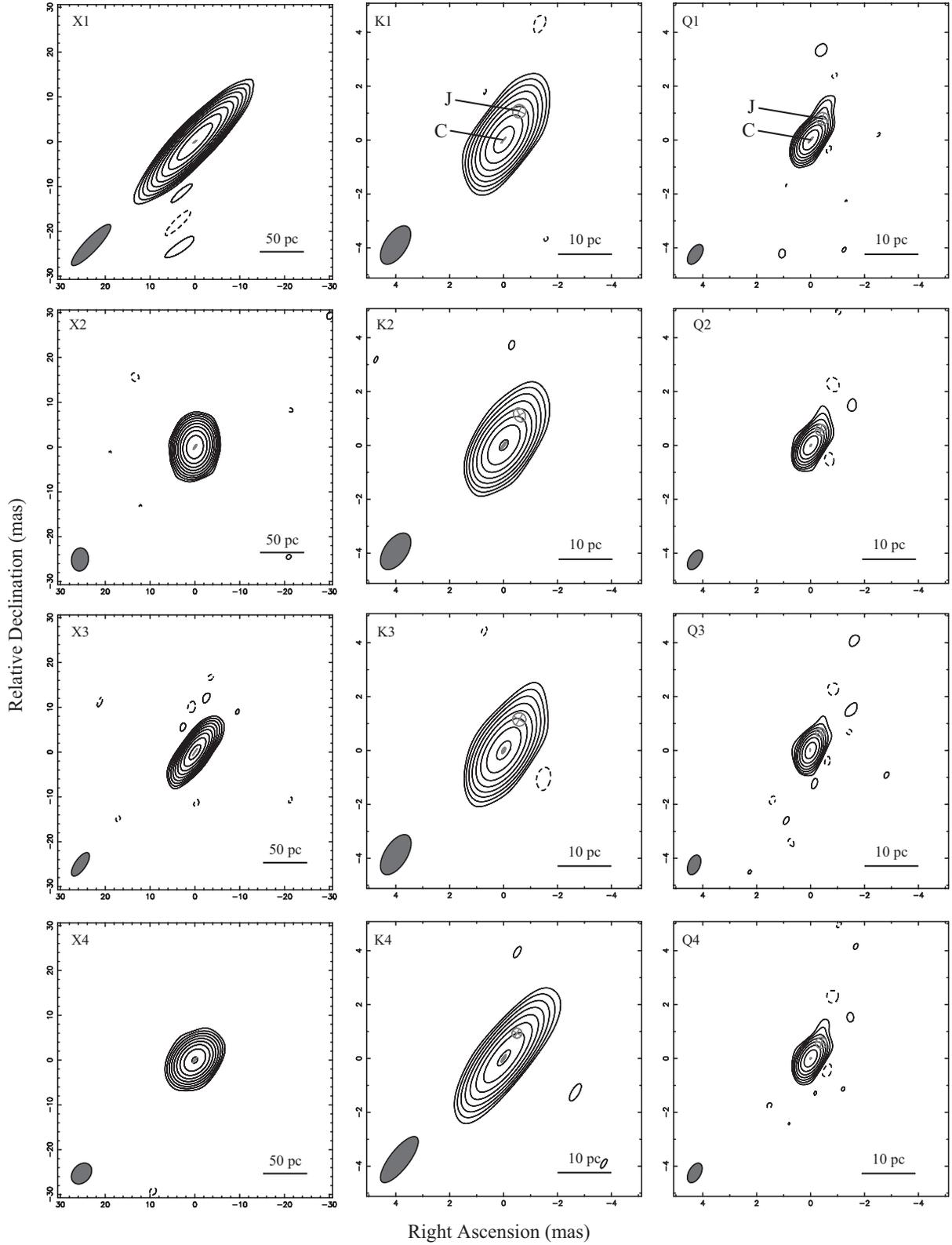}
  \end{center}
  \caption{VLBI images. Left panels show the 8.4~GHz images, center panels for 22~GHz, right panels for 43~GHz images, respectively. 
Images of 1st, 2nd, 3rd, and 4th observations are shown from the top to the bottom rows.
Contours indicate $-3\sigma, 3\sigma \times 2^n (n = 0, 1, 2, \cdots )$ of the intensity of each image. Synthesized beams are shown at left bottom of each image.
Elliptical and circular gaussian models are overlaid as the gray symbols. 
Due to the lack of Ogasawara of VERA, the beam was elongated in X1.}
  \label{fig:figure2}
\end{figure*}

The images obtained for each epoch at 8.4, 22 and 43~GHz are shown in figure \ref{fig:figure2}.
The beam size, peak intensity, image noise and dynamic range are listed in table \ref{tabl:table2}.
All VLBI images at 22 and 43~GHz show a compact core and a jet extended to northwest at position angle $\sim -30^\circ $.
These images are in agreement with that of the previous observations (e.g., \cite{2004ApJS..150..187O}; \cite{2005AJ....130.1418J}).
A point source structure was observed at 8.4~GHz at all epochs.
We applied the Gaussian model to the visibilities to derive flux density, position, and size of the components.
Single Gaussian model was used for 8.4~GHz.
For 22 and 43~GHz, elliptical Gaussian for the core and circular Gaussian for the jet, respectively, were applied.
The derived parameters of model-fitting,
the major and minor axes of elliptical Gaussian, and the diameter of circular Gaussian are listed in table \ref{tabl:table3}.
The models are also shown in figure \ref{fig:figure2}.
In the epoch Q3, the derived core model converged to a line (minor axis is zero), and only the major axis is listed in table \ref{tabl:table3}.

The average sizes of the core (geometric mean of major and minor axes, and arithmetic mean of them for all epochs) 
at 8.4~GHz are 0.46~mas = 2.3~pc, 0.14~mas = 0.70~pc at 22~GHz, and 0.073~mas = 0.37~pc (not includes the core size of Q3) at 43~GHz, respectively.
Since the observed core sizes of PKS~1510$-$089 are much smaller than the beam sizes,
those observed sizes are not confident but should be substantially considered as upper limits.
It seemed that the jet structure did not change in terms of its position, structure, and size within the error over four epochs at 22 and 43~GHz.
The weakness of the jet ($\leq 5~\% $ compare to the core) and the location of the jet close to the core do not allow us
to discuss more on the valiability of the jet quantitatively.

\begin{table*}[tbp]
\begin{center}
\caption{Parameters of VLBI images.}
\label{tabl:table2}
\begin{tabular}{cccrccr} \hline
Freq. & Epoch & \multicolumn{2}{c}{Synthesized beam} &  $S_\mathrm{peak}$ & Image noise & \multicolumn{1}{c}{DR} \\
(GHz) &        & \multicolumn{1}{c}{(mas$\times$mas)}  & \multicolumn{1}{c}{(deg)}  & (Jy/beam) & (mJy/beam) & \\\hline
8.4   & X1      &  12.4 $\times$ 2.97  & $-$43.4  & 1.86  & 1.41  & 1323 \\
      & X2      &  5.16 $\times$ 3.77  & $-$6.1   & 1.63  & 1.10  & 1482 \\
      & X3      &  6.13 $\times$ 2.42  & $-$34.6  & 1.65  & 1.65  & 1001 \\
      & X4      &  5.23 $\times$ 3.86  & $-$42.1  & 2.00  & 2.48  &  808 \\\hline
22    & K1      &  1.62 $\times$ 0.82  & $-$33.3  & 1.36  & 2.63  &  517 \\
      & K2      &  1.54 $\times$ 0.86  & $-$35.0  & 1.38  & 3.41  &  405 \\
      & K3      &  1.68 $\times$ 0.83  & $-$32.2  & 1.56  & 3.51  &  444 \\
      & K4      &  2.09 $\times$ 0.72  & $-$37.8  & 1.55  & 3.73  &  415 \\\hline
43    & Q1      &  0.83 $\times$ 0.44  & $-$32.9  & 1.09  & 2.99  &  364 \\
      & Q2      &  0.80 $\times$ 0.44  & $-$31.2  & 1.22  & 3.38  &  361 \\
      & Q3      &  0.77 $\times$ 0.44  & $-$22.3  & 1.36  & 4.43  &  307 \\
      & Q4      &  0.79 $\times$ 0.44  & $-$29.4  & 1.09  & 3.36  &  325 \\\hline
 \multicolumn{7}{@{}l@{}}{\hbox to 0pt{\parbox{120mm}{\footnotesize
Notes. 
Col. 1: observation frequency; 
Col. 2: observation epoch; 
Col. 3: synthesized beam size at the time of the analysis; 
Col. 4: position angle of the major axis of synthesized beam at the time of analysis; 
Col. 5: peak intensity of map; 
Col. 6: image noise 1$\sigma $; 
Col. 7: dynamic range, i.e. the ratio of $S_\mathrm{peak}$ and image noise.
 }\hss}}
\end{tabular}
\end{center}
\end{table*}

\begin{table*}[tbp]
\caption{Parameters of model fitting.}
\label{tabl:table3}
\begin{center}
\begin{tabular}{cccccrccr} \hline
Freq. & Epoch & Component & Flux & Separation & \multicolumn{1}{c}{P.A.} & \multicolumn{3}{c}{Size and P.A.} \\\cline{7-9}
& & & & & & Major & Minor & $\phi $ \\
(GHz) &  & & (Jy) & (mas) & \multicolumn{1}{c}{(deg)} & (mas) & (mas) & (deg) \\\hline
8.4 & X1 &  & 1.87 &   &   & 0.98 & 0.16 & $-$53.2 \\
 & X2 &  & 1.70 &   &   & 1.32 & 0.31 & $-$32.5 \\
 & X3 &  & 1.72 &   &   & 1.85 & 0.08 & $-$34.7 \\
 & X4 &  & 2.07 &   &   & 1.31 & 0.14 & $-$38.5 \\\hline
22 & K1 & C & 1.36 &   &   & 0.27 & 0.02 & $-$38.5 \\
 & K1 & J & 0.10 & 1.22  & $-$28.4  & 0.49 & &  \\
 & K2 & C & 1.41 &   &   & 0.34 & 0.07 & $-$36.1 \\
 & K2 & J & 0.09 & 1.27  & $-$27.5  & 0.51 &  &  \\
 & K3 & C & 1.56 &   &   & 0.24 & 0.11 & $-$26.9 \\
 & K3 & J & 0.07 & 1.29  & $-$26.8  & 0.49 &  &  \\
 & K4 & C & 1.54 &   &   & 0.33 & 0.08 & $-$32.2 \\
 & K4 & J & 0.08 & 1.06  & $-$27.8  & 0.36  & &  \\\hline
43 & Q1 & C & 1.16 &   &   & 0.27 & 0.04 & $-$40.3 \\
 & Q1 & J & 0.07 & 0.91  & $-$29.0  & 0.45  & &  \\
 & Q2 & C & 1.22 &   &   & 0.09 & 0.03 & $-$31.5 \\
 & Q2 & J & 0.06 & 0.69  & $-$33.8  & 0.41 & &  \\
 & Q3 & C & 1.38 &   &   & 0.09 &  & 10.7 \\
 & Q3 & J & 0.04 & 0.76  & $-$30.3  & 0.34 & & \\
 & Q4 & C & 1.10 &   &   & 0.06 & 0.03 & $-$59.0  \\
 & Q4 & J & 0.04 & 0.70  & $-$34.7  & 0.42 & &  \\\hline
 \multicolumn{9}{@{}l@{}}{\hbox to 0pt{\parbox{140mm}{\footnotesize
Notes. Col. 1: observation frequency; Col. 2: observation epoch; Col. 3: VLBI component, C: core; J: jet; 
Col. 4: Flux density of component (Jy); Col. 5: Separation from the map center (mas); 
Col. 6: Positon Angle which took north to 0$^\circ $ and made the counterclockwise rotation positive;
Col. 7-9: Gaussian model size and position angle of major axis. 
Major and Minor are the major axis and minor axis of elliptial Gaussian (FWHM), others are the diameter of circular Gaussian. 
$\phi $ is position angle which took north to 0 for major axis of elliptical Gaussian model and made the counterclockwise rotation positive.
 }\hss}}
\end{tabular}
\end{center}
\end{table*}

\subsection{Variability}
\begin{figure}
  \begin{center}
    \FigureFile(80mm,80mm){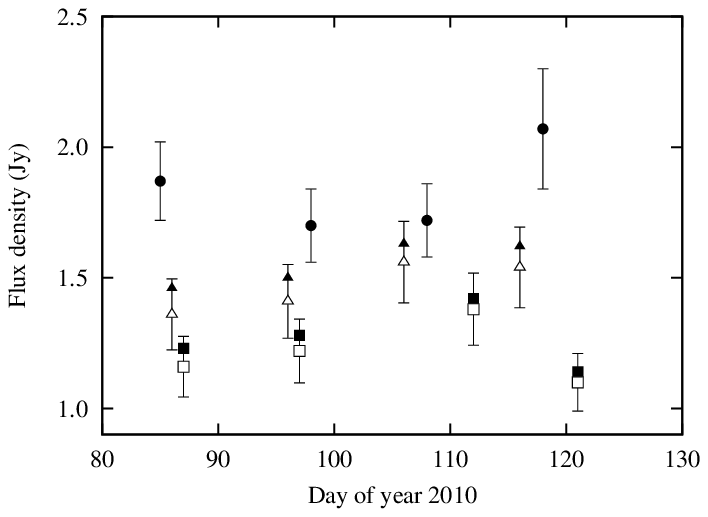}
  \end{center}
  \caption{Flux density variations of VLBI components. 
The symbols for 8.4, 22, 43~GHz are circle, triangle, square respectively.
Filled symbols represent the sum of core and jet components. Open symbols represent the core components.}
  \label{fig:figure3}
\end{figure}

Light curves of the VLBI components are shown in figure \ref{fig:figure3}.
Significant variability was observed at all three bands.
The light curves of 22~GHz (triangle) and 43~GHz (square) are similar each other.
They slightly increased between K1/Q1 and K2/Q2, and rapidly increased between K2/Q2 and K3/Q3. The flux of 43~GHz suddenly decreased at Q4.

\begin{figure}
  \begin{center}
    \FigureFile(80mm,80mm){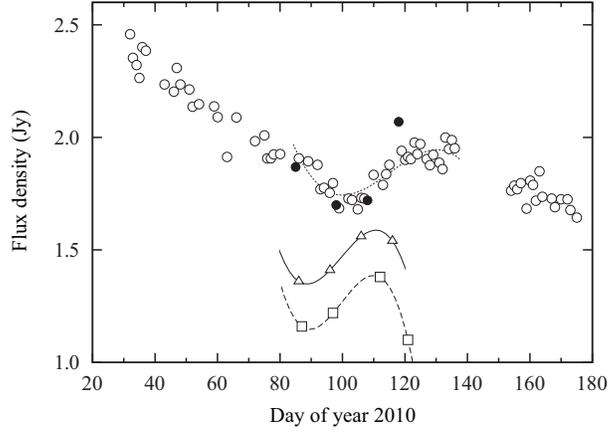}
  \end{center}
  \caption{VLBI and total flux density light curves of PKS~1510$-$089. 
  Open circle: total flux density at 8.4~GHz; filled circle: VLBI flux density at 8.4~GHz.
Open triangles and open squares are the same as those in figure 3.
Dotted line, solid line and dashed line indicate the best-fit third order polynomial for 8.4~GHz (total), 22 and 43~GHz (core).}
\label{fig:figure4}
\end{figure}

The VLBI core and total flux variabilities were overlaid in figure \ref{fig:figure4}.
The VLBI flux density of 8.4~GHz (filled circle) is in good agreement with the total flux density
(the 4th VLBI flux is higher than the total flux but within the 1 sigma VLBI flux error bar),
suggesting that the flux of diffuse emission of PKS~1510$-$089 is negligible at 8.4~GHz.
The variability observed by VLBI at 8.4~GHz corresponded to the variation around the minimum flux density observed by Yamaguchi 32~m at DOY 100.
The total flux density at 8.4~GHz reached the maximum at DOY 130, and decreased later.
The behavior of the variability at 8.4~GHz resembles to that at 22 and 43~GHz
because the flux densities at all band show minimum, increase, maximum, and decrease around DOY 90 to 130.

The variation pattern at 22 and 43~GHz preceded with that of 8.4~GHz.
We fit a third order polynomial to the light curves of the core at 22 and 43~GHz (solid line for 22~GHz and dashed line for 43~GHz in figure \ref{fig:figure4}),
and estimated the date and flux density of the minimum and the maximum as well as that at 8.4~GHz.
The date of minimum was DOY 89 and 90, the maximum was DOY 111 and 110 at 22 and 43~GHz, respectively.
The periods between the minimum and the maximum are 22 and 20 days at 22 and 43~GHz, respectively.
The change of the flux density of the core at 22~GHz and 43~GHz are substantially synchronized.
The dates and the flux densities at the minimum and the maximum were summarized in table \ref{tabl:table4}.

\begin{table}[tbp]
\caption{Flux density at the maximum and the minimum, and their date.}
\label{tabl:table4}
\begin{center}
\begin{tabular}{ccccc} \hline
Freq. & \multicolumn{2}{c}{Minimum} & \multicolumn{2}{c}{Maximum} \\
(GHz) & DOY & $S_\mathrm{min}$(Jy) & DOY & $S_\mathrm{max}$(Jy) \\\hline
8.4 & 100 & 1.74 & 130 & 1.95 \\
22\footnotemark[$\ast $] & 89 & 1.35 & 111 & 1.59 \\
43\footnotemark[$\ast $] & 90 & 1.15 & 110 & 1.39 \\ 
\hline
 \multicolumn{5}{@{}l@{}}{\hbox to 0pt{\parbox{80mm}{\footnotesize
\par\noindent
\footnotemark[$\ast $] VLBI core component.\\
 }\hss}}
\end{tabular}
\end{center}
\end{table}

\section{Discussion}
\subsection{Short-Term Variability and Delay between Frequencies}

Significant variations of the flux density were observed at all frequencies from 8.4 to 43~GHz.
The light curves showed the minimum and the maximum during the observation, and the periods
between the minimum and the maximum were 20 to 30 days. The variation patterns well alike at
3 frequencies, moreover, the variation at 22 and 43~GHz were closely synchronized. It is hard
to cause such a large variation ($\sim 20~\%$) by interstellar scintillation at frequency as high as 43~GHz.
We conclude that those short-term variabilities are intrinsic.

The variation was synchronized between 22 and 43~GHz, and the spectral index $\alpha$ ($S_\nu \propto \nu ^\alpha $) was almost constant at $-0.2$ to $-0.3$.
Since the spectral index does not change while the flux density varied,
the radio emitting region is optically thin at 22 and 43~GHz.
On the other hand, the variation at 8.4~GHz was delayed 11 to 20 days from that of 22 and 43~GHz.
When the small flare peaked around DOY 110 at 22 and 43~GHz, flux density at 8.4~GHz was still small.
The existence of a time lag in the variability suggests that the 8.4~GHz emitting region is optically thick.
The observed VLBI core size was 0.5 mas or less at 8.4~GHz, and the peak flux density was about 2 Jy.
These parameters suggested that the synchrotron self-absorption occurred at 8.4~GHz.

The variability behaviors of the three bands would be explained as follows (\cite{2006MNRAS.373.1470P}, \yearcite{2007MNRAS.381..797P}):
high-energy electrons which caused the small flare were injected in a blob at around DOY 110
and the flux density of the blob increased at 22 and 43~GHz, but did not change at 8.4~GHz
because the blob was optically thick at that frequency.
Then the high energy electrons loosed their energy while the emitting blob expanded around DOY 120
and the flux density decreased at 22 and 43~GHz, but increased at 8.4~GHz 
along with the increment of the solid angle of the blob.

\subsection{Doppler Factor and Brightness Temperature}
Many studies have been made on the derivation of Doppler factor for this source.
\citet{2010ApJ...721.1425A} obtained $\delta = 21$ from the superluminal velocity.
From the inverse Compton model, $\delta = 14.5$ or 11 (\cite{1993ApJ...407...65G}, \cite{1996ApJ...461..600G}),
from the equipartition model \citep{1994ApJ...426...51R}, $\delta = 10$ \citep{1996ApJ...461..600G},
and from the equipartition and variability, $\delta = 16.7$ \citep{2009A&A...494..527H},
were derived for PKS~1510$-$089.

The Doppler factor can be estimated from the time scale of variability $\Delta t$ as follows \citep{2005AJ....130.1418J},
\begin{equation}
	\delta _\mathrm{var} = \frac{sD_\mathrm{L}}{c \Delta t (1+z)}
\end{equation}
where $D_\mathrm{L} = 1906.9$~Mpc is the luminosity distance of PKS~1510$-$089, $s$ is the angular size of a core, i.e., 1.6 times the FWHM of the Gaussian model.
We adopted $\Delta t$ of 20~days as the period between the minimum and the maximum at 43~GHz, and $s = 0.073 \times 1.6$~mas,
and derived Doppler factor $\delta_\mathrm{var}$ of 47. Since the size is an upper limit, the Doppler factor is in fact an upper limit.
This Doppler factor is consistent with that \citet{2005AJ....130.1418J} derived $\delta_\mathrm{var} = 34.2-41.3$ for three components of PKS~1510$-$089 with VLBA at 43~GHz.
The Doppler factor of 47 is in the range of the observed distribution of $\delta$ by \citet{2004ApJ...609..539K}.
If the short-term variation observed in PKS~1510$-$089 is due to the Doppler effect,
other AGNs with relatively high Doppler factors would show the short-term variation like PKS~1510$-$089.

We estimated the observed brightness temperature in two ways:
$T_{b, \mathrm{size}}$ from the observed size $\theta$,
and $T_{b, \mathrm{var}}$ from the variability timescale $\Delta t$.
These two estimates of brightness temperature are derived from the following equations (2) and (3):
\begin{equation}
	T_{b, \mathrm{size}} = \frac{c^2}{2 k_\mathrm{B} \nu ^2} \frac{4}{\pi \theta ^2} S_{\rm max}
\end{equation}
\begin{equation}
	T_{b, \mathrm{var}} = \frac{2}{\pi k_\mathrm{B} \nu ^2} \left( \frac{D_\mathrm{L}}{\Delta t (1+z)} \right) ^2 \Delta S_{\nu },
\end{equation}
where $\Delta S_{\nu } = S_{\rm max} - S_{\rm min}$.
The derived brightness temperatures $T_{b, \mathrm{size}}$ and $T_{b, \mathrm{var}}$ are listed in table \ref{tabl:table5}.
The variability brightness temperatures are much higher than the size brightness temperatures.
This is because $T_{b, \mathrm{var}}$ is boosted by Doppler factor much larger than the boosting for $T_{b, \mathrm{size}}$.
These observed brightness temperatures are related to the intrinsic brightness temperature $T_{b, \mathrm{int}}$ as
$T_{b, \mathrm{size}} = \delta ^{1-\alpha }/(1+z)^{1-\alpha }~T_{b, \mathrm{int}}$ and 
$T_{b, \mathrm{var}} = \delta^{3-\alpha }/(1+z)^{3-\alpha }~T_{b, \mathrm{int}}$, respectively (\cite{2005AJ....130.1418J}, \cite{1999ApJ...511..112L}).
After the correction of Doppler boosting, the intrinsic brightness temperatures drived from $T_{b, \mathrm{size}}$ and $T_{b, \mathrm{var}}$
are also listed in table \ref{tabl:table5} (we assumed $\alpha = -0.3$).
The intrinsic brightness temperatures are in good agreement.

\begin{table}[tbp]
\caption{Brightness temperature.}
\label{tabl:table5}
\begin{center}
\begin{tabular}{cccccc} \hline
    & \multicolumn{2}{c}{Observed}  & \multicolumn{2}{c}{Intrinsic} \\ 
Freq.	& $T_{b, \mathrm{size}}$ & $T_{b, \mathrm{var}}$ & $T_{b, \mathrm{size}}$ & $T_{b, \mathrm{var}}$  \\
(GHz)	& ($10^{11}$~K) & ($10^{14}$~K) & ($10^{9}$~K) & ($10^{9}$~K)  \\\hline
8.4	& $\ge 2.3$	& $3.9$ & $\ge 2.3$ & $\ge 3.2$ \\
22	& $\ge 3.0$	& $1.2$ & $\ge 3.0$ & $\ge 1.0$  \\
43	& $\ge 2.5$	& $0.38$ & $\ge 2.5$ & $\ge 0.32$  \\
\hline
 \multicolumn{5}{@{}l@{}}{\hbox to 0pt{\parbox{80mm}{\footnotesize
\par\noindent
 }\hss}}
\end{tabular}
\end{center}
\end{table}

\section{Conclusion}
It is said that the time scale of flux variation in AGN is long at microwave compare to the other high frequency bands.
We studied whether there would be any short-term intrinsic variation with time scale less than 30 days at 8 to 43~GHz, 
and any correlation between the variability at different frequencies if such short-term variability existed.
We selected PKS~1510$-$089 as a target of observation, and performed flux measurement 72 times in 143 days at 8.4~GHz by
Yamaguchi 32~m radio telescope. Also VLBI observations for four times at 8.4, 22 and 43~GHz,
were performed every about ten days during the flux measurement. As a result, significant
variations of flux density were detected in all frequencies. The minimum and maximum were observed
during the observation, and the time interval of the minimum and maximum were 20 to 30 days. The variations were synchronized
between 22 and 43~GHz, while at 8.4~GHz the variation was similar shape but delayed 11 to 20 days. Since the patterns
of variation were well alike among 3 frequencies, this variation is not caused by scintillation but
an intrinsic one.
Change of spatial structure on a scale of $\sim$10~pc was not observed.
The Doppler factor calculated from the time scale and size of the core was 47, and is consistent
with the previous studies.
If the short-term variation observed in PKS~1510$-$089 is due to the Doppler effect,
other AGNs with relatively high Doppler factors would show such short-term intrinsic variability.

\bigskip
The authors thank the JVN team for observation assistance and support. 
The JVN project is led by the National Astronomical Observatory of Japan (NAOJ), 
which is a branch of the National Institutes of Natural Sciences (NINS), Hokkaido University, Ibaraki University, Tsukuba University, Gifu University, 
Osaka Prefecture University, Yamaguchi University, and Kagoshima University, in cooperation with the Geographical Survey Institute (GSI), 
the Japan Aerospace Exploration Agency (JAXA), and the National Institute of Information and Communications Technology (NICT).
The authors also thank to Dr. Hiroshi Nagai and Dr. Svetlana Jorstad for their helpful comments.

\end{document}